\documentclass[reprint, 
superscriptaddress,
 amsmath,amssymb,
 aps,
]{revtex4-1}

\usepackage{graphicx}
\usepackage{dcolumn}
\usepackage{bm}
\usepackage{hyperref}
\usepackage{upgreek}
\usepackage{xcolor}
\usepackage{mathtools}

\begin{document}

\title{Theory of Cation Solvation in the Helmholtz Layer of Li-ion Battery Electrolytes}

\author{Zachary A. H. Goodwin}
\email{zac.goodwin@materials.ox.ac.uk}
\affiliation{Department of Materials, University of Oxford, Parks Road, Oxford OX1 3PH, United Kingdom}
\affiliation{John A. Paulson School of Engineering and Applied Sciences, Harvard University, Cambridge, Massachusetts 02138, United States}

\author{Daniel M. Markiewitz}
\affiliation{Department of Chemical Engineering, Massachusetts Institute of Technology, Cambridge, Massachusetts  02139, USA}

\author{Qisheng Wu}
\affiliation{School of Engineering, Brown University, Providence, RI, 02912, USA}

\author{Yue Qi}
\affiliation{School of Engineering, Brown University, Providence, RI, 02912, USA}


\author{Martin Z. Bazant}
\affiliation{Department of Chemical Engineering, Massachusetts Institute of Technology, Cambridge, Massachusetts  02139, USA}
\affiliation{Department of Mathematics, Massachusetts Institute of Technology, Cambridge, Massachusetts 02139, USA}

 
\date{\today}

\begin{abstract}
    The solvation environments of Li$^+$ in conventional non-aqueous battery electrolytes, such as LiPF$_6$ in mixtures of ethylene carbaronate (EC) and ethyl methyl carbonate (EMC), are often used to rationalize the transport properties of electrolytes and solid electrolyte interphase (SEI) formation. In the SEI, the solvation environments in the compact electrical double layer (EDL) next to the electrode, also known as the Helmholtz layer, determine (partially) what species can react to form the SEI, with bulk solvation environments often being used as a proxy. Here we develop and test a  theory of cation solvation in the Helmholtz layer of non-aqueous Li-ion battery electrolytes. First, we validate the theory against bulk and diffuse EDL atomistic molecular dynamics (MD) simulations of LiPF$_6$ EC/EMC mixtures as a function of surface charge, where we find the theory can capture the solvation environments well. Next we turn to the Helmholtz layer, where we find that the main effect of the solvation structures next to the electrode is an apparent reduction in the number of binding sites between Li$^+$ and the solvents, again where we find good agreement with our developed theory. Finally, by solving a simplified version of the theory, we find that the probability of Li$^+$ binding to each solvent remains equal to the bulk probability, suggesting that the bulk solvation environments are a reasonable place to start when understanding new battery electrolytes. Our developed formalism can be parameterized from bulk MD simulations and used to predict the solvation environments in the Helmholtz layer, which can be used to determine what could react and form the SEI.
\end{abstract}

\maketitle

\section{Introduction}

Lithium-ion batteries are set to play a central role in our efforts to de-carbonize transportation and the storage of locally produced renewable energy, which will aid our efforts against curbing global warming~\cite{Tian2021,li2020new,li2020high,wang2022liquid}. One of the central components of a Li-ion battery is the liquid electrolyte that transports the Li$^+$ between the cathode and anode to store/release energy~\cite{Meng2022,li2020new}. The electrolytes that are used typically contain fluorinated anions, such as PF$_6^-$, and carbonate-based solvents, such as ethylene carbonate (EC) and ethyl methyl carbonate (EMC), with a salt concentration of  $\sim$1~M~\cite{xu2004nonaqueous,xu2014electrolytes,yu2020molecular,yu2022rational}. The carbonate solvents strongly interact with and solvate the Li$^+$ ions through the carbonyl functional group, which regulates ionic aggregates at this relatively high salt concentration, and therefore, ensures good transport properties~\cite{xu2007solvation,von2012correlating,Li2015,Piao2022,Wu2023,Goodwin2023PRXE}. One of the key observations in the field of battery electrolytes is the link between the solvation environments of active cations, and the physical properties, such as conductivity, transference numbers and formation of the solid electrolyte interphase (SEI), which is linked to the long-term cycling ability of batteries~\cite{xu2007solvation,Borodin2014SEI,Zheng2017Uni,Piao2022,Cheng2022Sol,Borodin2014SEI,Wu2023,borodin2020uncharted}. As bulk solvation environments are readily accessible from experiments~\cite{barthel2000ftir,Cresce2017,seo2012electrolyte1,seo2012electrolyte2,yang2010investigation,von2012correlating,Zhang2018ISI} and simulations~\cite{xu2007solvation,Postupna2011,von2012correlating,Borodin2014SEI,Li2015,Skarmoutsos2015,Borodin2017Mod,Han2017MD,Ravikumar2018,Shim2018,Han2019MD,Piao2020Count,Hou2021,Wu2022,Cresce2017,Piao2022,Wu2023,Goodwin2023PRXE,Efaw2023,Hossain2023}, these are often used as a starting point to understand battery electrolytes~\cite{chen2020electrolyte}.


However, what reacts at the electrode and forms the SEI is linked to the composition of the electrolyte at the charged interface, which generally is not the same as the bulk composition~\cite{von2012correlating,Wu2023,Oyakhire2023,Wu25}. More generally, without reactions, this is known as the electrical double layer (EDL) of the electrolyte~\cite{Fedorov2014,Bazant2009a,goodwin2017mean}, with the electrolyte directly in contact with the electrode often being referred to as the Helmholtz layer (or Stern layer), and the diffuse EDL is the distribution of the electrolyte which screens the remaining charge of the electrode, as depicted in Fig.~\ref{fig:schematic}. In the context of conventional non-aqueous battery electrolytes, a large body of literature exists on simulating the EDL with atomistic methods, such as classical molecular dynamics (MD) and \textit{ab initio} MD, where changes in composition of the electrolyte and solvation environments have been rationalized and used to interpret SEI formation~\cite{von2012correlating,Wu2023,Oyakhire2023}. This area is further burgeoning with the promise of reactive force fields using machine learning interatomic potentials~\cite{Yao2022CRev,Ren24,Ioan2022,Yang24,Goodwin24MLIL}, and reaction networks~\cite{Xie21,Spotte22} which have given great insight into SEI formation so far.


The EDL of electrolytes also has a long history of being studied with relatively simple thermodynamic theories~\cite{Fedorov2014,Bazant2009a}. In the context of battery electrolytes, however, this area appears to be less well developed, as the important solvation structures are often not accounted for with simple electrolyte theories~\cite{Goodwin2023PRXE}. Recently, McEldrew and Goodwin \textit{et al.}~\cite{mceldrew2020theory,mceldrew2020corr,mceldrew2021ion,mceldrew2020salt,Goodwin2022GEL,Goodwin2022IP,Goodwin2023PRXE} have applied the reversible polymerisation theories of Flory~\cite{flory1941molecular,flory1941molecular2,flory1942constitution,flory1942thermodynamics,flory1953principles,flory1956statistical}, Stockmayer~\cite{stockmayer1943theory,stockmayer1944theory,stockmayer1952molecular} and Tanaka~\cite{tanaka1989,tanaka1990thermodynamic,tanaka1994,ishida1997,tanaka1995,tanaka1998,tanaka2002,tanaka1999,tanaka2011polymer} to concentrated electrolytes, where ionic aggregation and solvation have been rationalized with a simple, analytical theory. Moreover, Markiewitz \textit{et al.} have recently extended this theory to the EDL of several realistic electrolytes~\cite{Markiewitz25WiSE,Zhang24,Markiewitz24}. However, Markiewitz \textit{et al.}~\cite{Markiewitz25WiSE} found that the largest deviation between their theory and MD simulation occurred right at the interface, i.e., in the Helmholtz layer. Therefore, further development of this theory for the Helmholtz layer is needed~\cite{Goodwin2022GEL}, and Li-ion battery electrolytes are an interesting system to start with because there are significant implications and applications for SEI formation.

In this paper, we develop and test a simple theory for the composition of the Helmholtz (or compact) double layer in conventional, non-aqueous Li-ion battery electrolyte mixtures. This theory is motivated from observations made from further analyzing the MD simulations performed by Wu \textit{et al.} in Ref.~\citenum{Wu2023}, where we find the main effect on the solvation structure in the Helmholtz layer is to reduce the number of available binding sites of Li$^+$, i.e., the surface blocks/binds to one or more of the available solvation sites of Li$^+$. %
First, we validate the bulk and diffuse EDL solvation environments against our theory, which work well, as shown in previous work, before moving onto the Helmholtz layer. By solving a simplified version of the theory in the Helmholtz layer, we find that the probability of Li$^+$ binding to the solvents remains constant and equal to the bulk value, at least in the assumptions of this simplified theory. Therefore, we find some theoretical foundation as to why studying the bulk solvation environments is a reasonable starting point for Li-ion battery electrolytes. 

\section{Methods}

\begin{figure}
    \centering
    \includegraphics[width=0.95\linewidth]{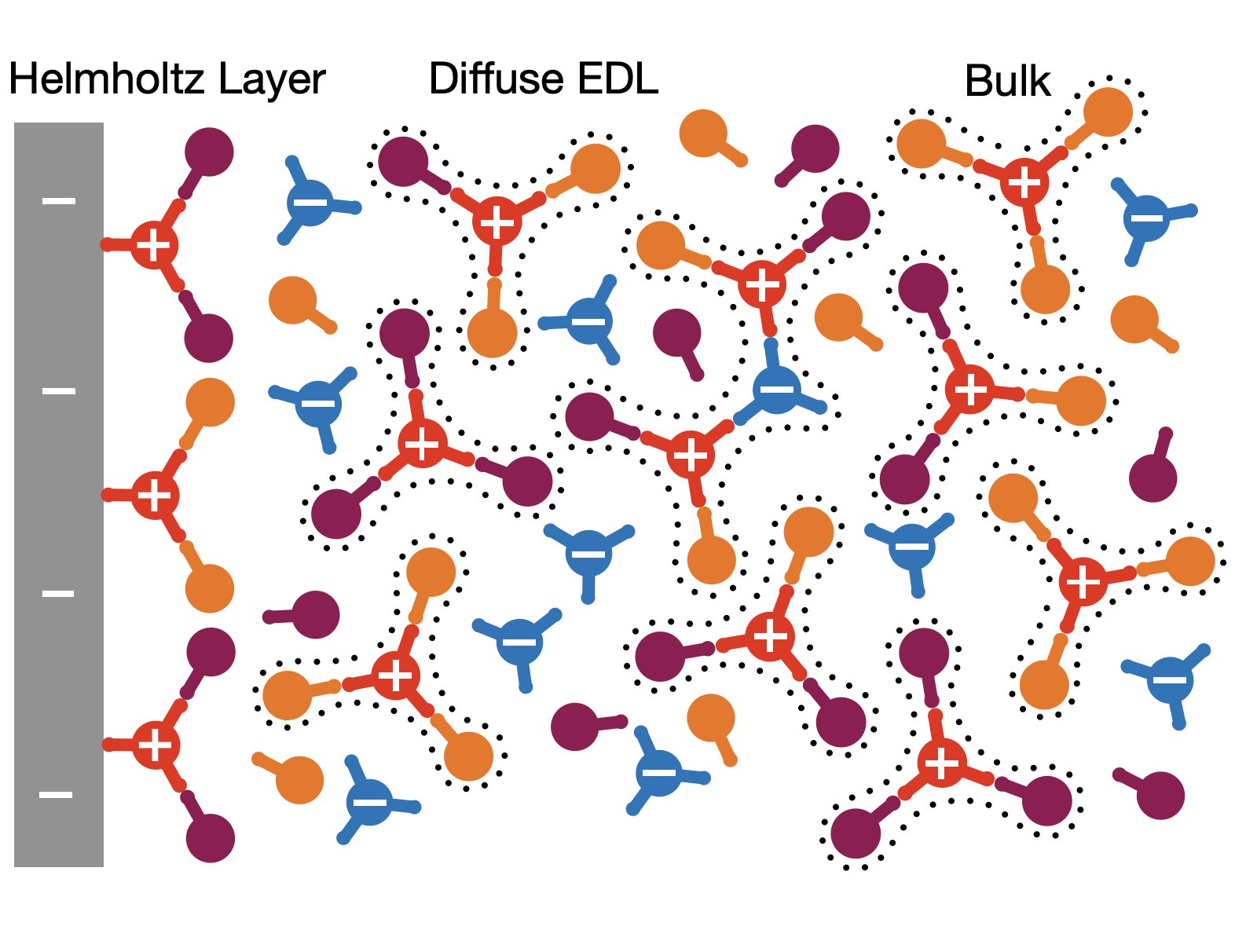}
    \caption{Schematic of conventional non-aqueous battery electrolytes in the bulk, where cation-solvation environments are depicted, in the diffuse electrical double layer (EDL), where larger aggregates are shown, and finally close to the interface there is the Helmholtz layer, where we have shown the cations interacting directly with the surface. One of the main parameters of the developed theory is the functionality of each species, i.e., the maximum number of associations it can form with other species. These functionalities are indicated as the sticks coming out of the circles for each species. For cations (denoted by at $+$) we have shown a functionality of 3, for anions (denoted with a $-$) we have again used 3, and the two solvents (distinguished by different colours) have a functionality of 1. We assume cation-solvent and cation-anion interactions are the only ones which dominate. At the Helmholtz layer, we find the interface blocks/binds to at least one of the cation association sites.}
    \label{fig:schematic}
\end{figure}


Here we further analyze the molecular dynamics simulations of several conventional battery electrolytes investigated in by Wu \textit{et al.}\cite{Wu2023}. Therefore, we refer the readers to Ref.~\citenum{Wu2023} for the details of the MD simulations. Here those EDL simulations are further analyzed in 3 sections: bulk, diffuse EDL and Helmholtz layer. The bulk region as defined as the middle 20~\AA~region (the distance between the two electrodes was set to around 100~\AA), the diffuse EDL is defined as from 5~\AA~from the interface to 10~\AA~from the interface, and the Helmholtz layer is defined from species at the interface to 5~\AA~(since this is the first layer of electrolyte in contact with the interface), as depicted in Fig.~\ref{fig:schematic}. Within each of these regions, we extract the numbers of each species, and define an association between Li$^+$ and F in PF$_6$ from a real-space cutoff of 2.8~\AA, and Li$^+$ and O (carbonyl) in different solvents from a real-space cutoff of 2.8~\AA, as explained in Ref.~\citenum{Wu2023}. These definitions of associations are then used to compute coordination environments and the number of each aggregate in each region. More details of this theoretical framework can be found in Refs.~\citenum{mceldrew2021ion,Goodwin2023PRXE}. Computing these associations allows us to investigate the essential components that a theory must have to be able to describe solvation in these different regions.


In this paper, we compare the MD determined cluster/solvent distribution against our theory in these different regions. As the bulk and diffuse EDL theory has been presented elsewhere, we refer the readers to Refs.~\citenum{mceldrew2020theory,mceldrew2020corr,mceldrew2021ion,mceldrew2020salt,Goodwin2022GEL,Goodwin2022IP,Goodwin2023PRXE,Markiewitz25WiSE,Zhang24,Markiewitz24} and the Supplementary Information (SI) for further details, and we will only provide an overview of the necessary equations and assumptions of the theory here. The central quantity that we are computing is the cluster/solvent distribution, as seen by
\begin{equation}
     c_{lmsq}=\frac{W_{lmsq}}{\lambda_{-}}\left(\psi_{l}\lambda_{-} \right)^{l} \left(\psi_{m}\lambda_{-} \right)^{m}(\psi_{s}\lambda_{x})^s(\psi_{q}\lambda_{y})^q.
    \label{eq:cluster_dist}
\end{equation}

\noindent Here $c_{lmsq}$ is the dimensionless concentration of a cluster of rank $lmsq$, which means there are $l$ cations, $m$ anions, $s$ solvent molecules of the first type ($x$) and $q$ solvent molecules of the second type ($y$) bound together in an aggregate. The dimensionless concentration is determined from $c_{lmsq} = N_{lmsq}/\Omega$, where $N_{lmsq}$ is the number of clusters of that rank and 
\begin{equation}
    \Omega = \sum_{lmsq}(l+\xi_-m+\xi_xs+\xi_yq)N_{lmsq}
\end{equation}

\noindent is the number of lattice sites occupied by the aggregates, where a single lattice site is set to the volume of the Li$^+$ cation ($v_+$), with $\xi_j = v_j/v_+$ being the volume ratio of each species to the Li$^+$ cation. Dividing through by the total number of lattice sites gives
\begin{equation}
    1 = \sum_{lmsq}(l+\xi_-m+\xi_xs+\xi_yq)c_{lmsq} = \sum_{lmsq}\phi_{lmsq}.
\end{equation}

\noindent which is a statement of incompressibility in the theory, where $\phi_{lmsq}$ is the volume fraction of a cluster of rank $lmsq$. It is also useful to know that the volume fraction of each species is determined through
\begin{equation}
    \phi_i = \sum_{lmsq}\xi_i j c_{lmsq},
\end{equation}

\noindent with the number of each species being determined from 
\begin{equation}
    N_i = \sum_{lmsq}j N_{lmsq}.
\end{equation}

In our theory, we assume that Cayley-tree like aggregates form, which means no loops can form, i.e., all of the aggregates are branched, as seen in Fig.~\ref{fig:schematic}. This is to ensure an analytically tractable theory, as the free energy of the associations can be uniquely determined from the number of species in the aggregates, and the configurational entropy is also uniquely determined. To form these Cayley-tree aggregates, we have to assume some maximum number of associations that the species can form, which we refer to as the functionality of the species, $f_i$. For cations and anions it is kept general ($f_+$ and $f_-$, respectively), but for solvent we assume that only 1 association with the cation may form (no anion-solvent solvation). This is particularly reasonable as in Li-salt electrolytes, the cation is small and binds with other species strongly, while the anion and solvent interactions are more comparable~\cite{Goodwin2023PRXE,Hou2021}. These assumptions have been verified for conventional battery electrolytes, and other electrolytes~\cite{Wu2023,Goodwin2023PRXE,Hou2021}. 

In Eq.~\eqref{eq:cluster_dist}, the next term in the equation is $W_{lmsq}$, given by
\begin{equation}
    W_{lmsq} = \dfrac{(f_+l -l)!(f_-m - m)!}{l!m!s!q!(f_+l - l - m - s - q +1)!}.
\end{equation}

\noindent which is related to the number of ways of arranging an aggregate of rank $lmsq$. The $\lambda_i$'s in Eq.~\eqref{eq:cluster_dist} are the association constants, as seen by
\begin{align}
    \lambda_{i}=e^{-\beta\Delta f_{+i}}
\end{align}

\noindent where $\Delta f_{+i}$ is the free energy of formation of an association between cations and $i$, with the reference state being the free species in solution~\cite{mceldrew2020theory}, and $\beta$ is the inverse thermal energy. Finally, $\psi_i = f_i\phi_{i}\alpha_i/\xi_i$ is the number of free association sites per lattice site for that species, where $\alpha_i$ is the fraction of that species $i = +,-,x,y$ that is free. 

The problem we now face is that we wanted to determine $\alpha_i$ from our theory, not have it be an input for the theory. To overcome this, we follow Tanaka~\cite{tanaka1989,tanaka1990thermodynamic,tanaka1994,ishida1997,tanaka1995,tanaka1998,tanaka2002,tanaka1999,tanaka2011polymer} and introduce association probabilities and their corresponding mass-action laws. Therefore, we introduce $\alpha_i = (1 - \sum_{i'}p_{ii'})^{f_i}$, where $p_{ii'}$ is the probability that $i$ is associated with $i'$. These probabilities are related through the conservation of associations 
\begin{align}
\psi_+ p_{+i} = \psi_i p_{i+} = \Gamma_{i},
\label{eq:sys1}
\end{align}

\noindent where $\Gamma_i$ is the number of $+i$ associations per lattice site, and the mass action laws
\begin{align}
    \lambda_{i}\Gamma_{i} = \frac{p_{i+}p_{+i}}{(1-\sum_{i'}p_{+i'})(1-p_{i+})}.
    \label{eq:sys2}
\end{align}

\noindent From solving this system of equations, the cluster distribution can be computed from the theory. All that is needed is the number of each species, $N_i$, the assumed functionalities for each species, $f_i$ and the volume ratios, $\xi_i$ (these are known from electrolyte composition), and to determine the association constant's, $\lambda_i$. Fortunately, the $\lambda_i$'s can be determined from the MD simulations~\cite{mceldrew2021ion,mceldrew2020corr,mceldrew2020salt}. First, the ensemble average coordination numbers of species associating to the cation are determined, which can then be divided by the cation functionality to find the association probabilities. The conservation of associations and mass action laws are then used to find the association constants. Therefore, there are no free fitting parameters of the theory, and we can investigate how the associations are behaving in the bulk, diffuse EDL and Helmholtz layer, to then inform a theory for the latter. 

Note that Goodwin \textit{et al.}~\cite{Goodwin2022GEL} showed that the same form of the cluster distribution should hold in the diffuse EDL, but where the quantities are replaced by their EDL counterparts, which is indicated with a bar. This theory was extended by Markiewitz \textit{et al.}~\cite{Markiewitz24,Markiewitz25WiSE,Zhang24} to deal with WiSE, and in general more realistic electrolytes. In the SI we again show the Hemlmotlz layer should also follow this cluster distribution, but where the volume fractions, association probabilities and association constants can be different from the bulk/diffuse EDL. Here we compare the theory and MD simulations through computing the $\lambda_i$ using the  $N_i$'s in the different regions for the most direct comparison. This means good agreement should be expected, but this allows us to verify the underlying assumptions of the theory and discover any new assumptions required for the Helmholtz layer.

From this full version of the theory, several assumptions can be made to investigate to a simpler set of equations. Firstly, as the Li-PF$_6$ associations are often weak, these can be neglected, i.e., $\lambda_-=0$ or from removing the association probabilities from the equations~\cite{Goodwin2023PRXE}. This allows one to focus on the solvation properties instead of aggregation effects. Secondly, the sticky-cation assumption can be employed, where $1 = p_{+x} + p_{+y}$. The reader is referred to Refs.~\citenum{Goodwin2023PRXE} for more details.

\begin{figure*}
    \centering
    \includegraphics[width=0.8\linewidth]{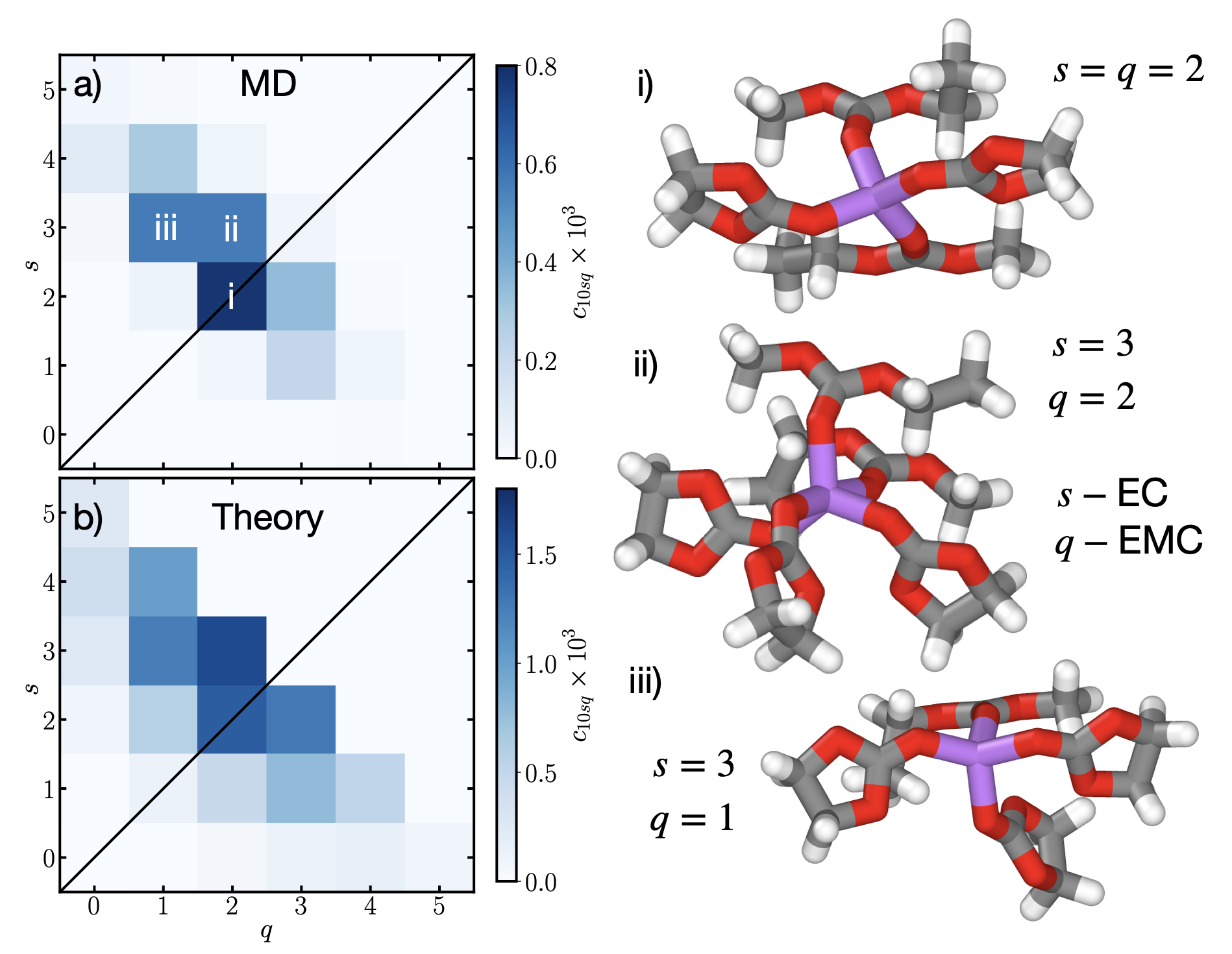}
    \caption{Solvation distributions, $c_{10sq}$, of Li$^+$ in the bulk from MD a) and theory b) as a function of the number of coordinating EC ($s$) and EMC ($q$). In i), ii) and iii), example solvation environments for 2EC$+$2EMC, 3EC$+$2EMC and 3EC$+$EMC are, respectively, shown, which are the most common solvation environments in MD simulations, as also indicated in a). These structures were visualized using Ovito~\cite{stukowski2009visualization}.}
    \label{fig:bulk}
\end{figure*}

\section{Results}

Here we show results for the 1~M LiPF$_6$ in EC-EMC 3:7 volume ratio electrolyte. In the SI we show equivalent results for the other electrolytes investigated in Ref.~\citenum{Wu2023}. Moreover, in the main text, we will only focus on the solvation properties of Li$^+$, not focusing on any ionic aggregation effects. In the SI, we show additional results for the comparison of the ionic aggregation in the bulk and diffuse EDL.

\subsection{Bulk}

In Fig.~\ref{fig:bulk}a) we show $c_{10sq}$, the concentration of the different solvation environments of Li$^+$, as a function of the number of solvating EC ($s$) and EMC ($q$) from the MD simulations, with the average number of solvents coordinated to Li$^+$ being 4.27. As seen, the most probable solvation structure is with 2 EC and 2 EMC solvating Li$^+$. The next most probable solvation environments are found to be 3 EC, and 1-2 EMC. We also find that there is some probability of solvation environments containing 4 EC and EMC, and 2 EC and 3 EMC. Overall, there is practically no example of $s+q > 5$, and typically no solvation environment with $s+q < 4$.

\begin{figure*}
    \centering
    \includegraphics[width=0.7\linewidth]{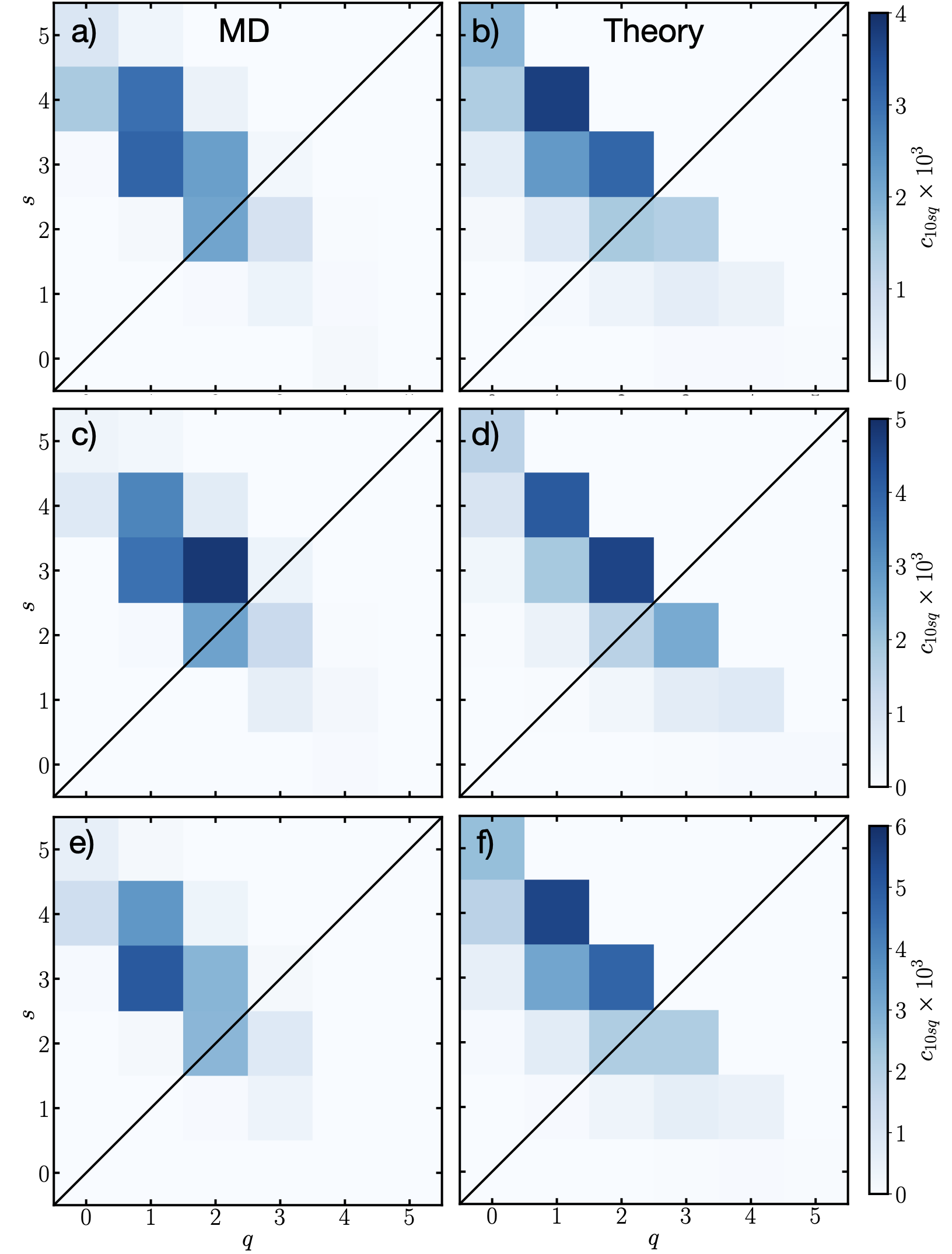}
    \caption{Solvation distributions, $c_{10sq}$, of Li$^+$ in the diffuse EDL from MD [a),c),e)] and theory [b),d),f)] as a function of the number of coordinating EC ($s$) and EMC ($q$) at, respectively, surface charges of $-0.4$, $-0.6$ and $-0.8$ enm$^{-2}$.}
    \label{fig:diffuse}
\end{figure*}

From these observations, a good choice for $f_+$ is $5$, with $4$ also being reasonable. In the context of WiSE, where there are similar average coordination numbers, it has been found that using 4 can result in better results~\cite{mceldrew2021ion,Markiewitz25WiSE}, but the sticky-cation case must then be used. As such, we choose to explore the case of $f_+ = 5$ here. In Fig.~\ref{fig:bulk}b), the theory cluster distribution is shown, plotted using Eq.~\eqref{eq:cluster_dist} and the association constants $\lambda_{x} = 136.96$  and $\lambda_{y} = 35.58$ (which were determined from the MD simulations using $f_+ = 5$, not accounting for any ionic associations). As can be seen, the most probable solvation structure involves 3EC and 2EMC, with the next most probable solvation environments containing 2EC$+$2EMC, 2EC$+$3EMC, 3EC$+$EMC, and 4EC$+$EMC. While the relative probabilities of these solvation environments do not exactly match the MD simulations, and moreover, the absolute values are slightly different, the overall trend of more EC in the solvation shell compared to EMC is captured. Despite there being more EMC in the electrolyte, the $\times 3$ larger association constant between Li-EC compared to Li-EMC results in EC slightly dominating the solvation shell. The disagreement between MD-Theory could be for a number of reasons, such as the MD simulation ensemble averages not being completely converged, or some assumptions of the theory breaking down, such as loop formation or higher-order interactions.

\subsection{Diffuse EDL}

Next, we turn to studying how the solvation environments change within the diffuse EDL, which is considered to be not the first 5~\AA~from the interface, but the next 5~\AA. In Fig.~\ref{fig:diffuse} the left column shows the MD results for $c_{10sq}$ and the right column shows the corresponding theory (calculated with the same method as the bulk). Each row in Fig.~\ref{fig:diffuse} is a different surface charge, starting from $-0.4$~e~nm$^{-2}$ in the top row, to $-0.6$~e~nm$^{-2}$ in the middle, to $-0.8$~e~nm$^{-2}$ in the bottom row. 

\begin{table}[]
    \centering
    \begin{tabular}{cccc}
    \hline 
    $\sigma /$ Cm$^{-2}$ & $\lambda_x/\lambda_y$ & $x_x/x_y$ & $p_{+x}$ \\    
    \hline \
        -0.4 & 3.07 & 1.41 & 0.64 \\
        -0.6 & 4.40 & 1.00 & 0.60 \\
        -0.8 & 6.50 & 1.50 & 0.63 \\
    \hline
    \end{tabular}
    \caption{Summary of association constant ratios and mole fraction ratios for the EC and EMC solvents for the diffuse EDL at the indicated surface charges.}
    \label{tab:diffuse}
\end{table}

For the MD results at $-0.4$~e~nm$^{-2}$, as seen in Fig.~\ref{fig:diffuse}a), the solvation structures are fairly similar to the bulk, albeit with larger concentrations of Li$^+$ solvation environments owing to the reduced anion concentration. The most probable solvation environment is 3EC$+$EMC, with 4EC$+$EMC being the next most probable. The corresponding theory calculation for $c_{10sq}$, using the association constants calculated from MD, displayed in Tab.~\ref{tab:diffuse}, is seen in Fig.~\ref{fig:diffuse}b). Clearly, there is a reasonable qualitative match with the MD, even though the exact ordering of the most probable solvation environments are not identical. The theory predicts 4EC$+$EMC to be the most likely, with 3EC$+$2EMC the next most probable.

At the more negative surface charge of $-0.6$~e~nm$^{-2}$, displayed in Fig.~\ref{fig:diffuse}c), $c_{10sq}$ is again relatively similar to the bulk. In this case, the most probable solvation structure is 3EC$+$2EMC. The theory calculation is shown in Fig.~\ref{fig:diffuse}d), where it also predicts that 3EC$+$2EMC is the most probable solvation environment, and it also predicts a similar distribution of solvation environments.

Finally, for a surface charge of $-0.8$~e~nm$^{-2}$, shown in Fig.~\ref{fig:diffuse}e) for MD, the solvation distribution is again relatively similar to the bulk. In this case the most probable solvation environment is 3EC$+$EMC, with 4EC$+$EMC being likely too. In Fig.~\ref{fig:diffuse}f) the corresponding theory is shown, where we find the most probable solvation environment to be 4EC$+$EMC. Again, there is reasonable agreement for the spread of solvation environments.

Overall, the solvation environments in the diffuse EDL are similar to those in the bulk, with the theory matching reasonably well against the MD simulations with $f_+ = 5$. In the SI the case of $f_+ = 4$ is shown, where slightly worse agreement is found. As seen in Tab.~\ref{tab:diffuse}, the ratio of the association constants and molar ratio is displayed for each surface charge. With more negative surface charge, $\lambda_x/\lambda_y$ increases slightly over the bulk value of 3.85, although not substantially. 
Moreover, the molar ratio of EC relative to EMC is now increased over the bulk value of $\sim$0.65, reflecting its preferred interaction with the electrostatic fields because of its larger dipole moment, but the association probability between Li$^+$-EC is practically constant.


\subsection{Helmholtz Layer}

\begin{figure*}
    \centering
    \includegraphics[width=0.7\linewidth]{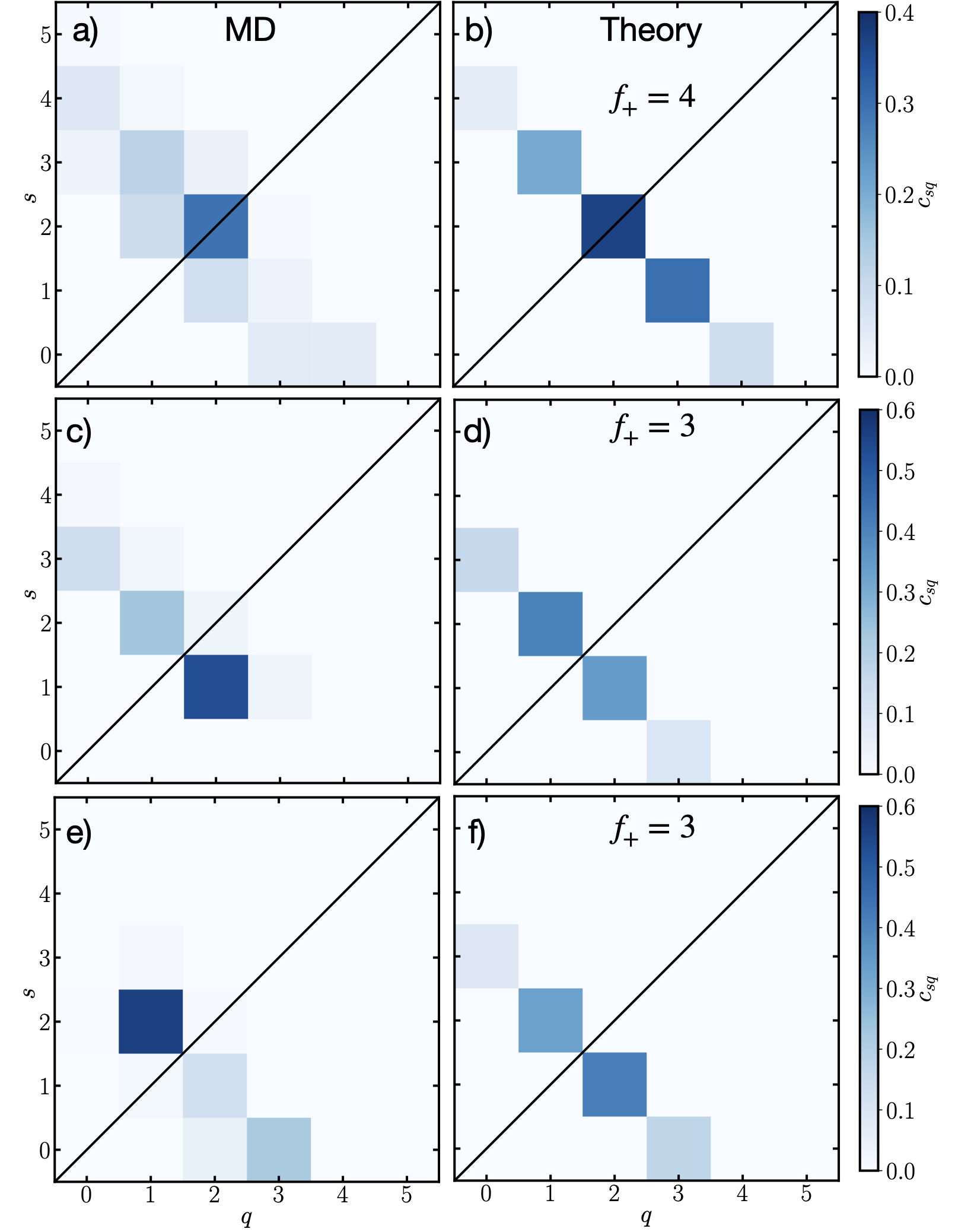}
    \caption{Solvation distributions, $c_{10sq}$, of Li$^+$ in the Helmholtz layer from MD [a),c),e)] and theory [b),d),f)] as a function of the number of coordinating EC ($s$) and EMC ($q$) at, respectively, surface charges of $-0.4$, $-0.6$ and $-0.8$ enm$^{-2}$.}
    \label{fig:sticky_helm}
\end{figure*}

Having demonstrated that the theory works well in the bulk and diffuse EDL, as previously found for other electrolytes~\cite{Markiewitz25WiSE}, in this section we turn to investigate the solvation environments of Li$^+$ in the Helmholtz layer of the anode, which corresponds to the first 5~\AA~next to the interface. In Fig.~\ref{fig:sticky_helm}a) we show the analysis for the Helmholtz layer for a surface charge of $-0.4$~e~nm$^{-2}$ from MD simulations. Similar to the bulk, we find that the most probable environment is 2EC$+$2EMC. However, there is practically no solvation structures with $s+q > 4$, and very little with $s+q < 4$. This is in contrast to the bulk case, when there were significant 5-coordinated Li, and a larger distribution of $s+q$. 


\begin{table}[]
    \centering
    \begin{tabular}{cccc}
    \hline
        $\sigma /$ Cm$^{-2}$ & $\lambda_x/\lambda_y$ & $x_x/x_y$ & $p_{+x}$ \\
    \hline
        -0.4 & 0.13 & 2.75 & 0.50 \\
        -0.6 & 0.11 & 4.21 & 0.52 \\
        -0.8 & 4.44$\times$10$^{-4}$ & 4.24 & 0.45\\
    \hline
    \end{tabular}
    \caption{Summary of association constant ratios and mole fraction ratios for the EC and EMC solvents for the Helmholtz layer at the indicated surface charges.}
    \label{tab:HLQ}
\end{table}


Therefore, it appears that the solvation environment of Li$^+$ is behaving in a sticky-way in the Helmholtz layer, which motivates us to compare the sticky-solvation theory against the MD simulations. Using $1 = p_{+x} + p_{+y}$ (normalized in MD such that this is true) and $f_+ = s + q$, we can arrive at
\begin{equation}
    \bar{c}_{sq} = \dfrac{\bar{c}_{10sq}}{\bar{\phi}_+} = \dfrac{f_+!}{s!(f_+ - s)!}\bar{p}_{+x}^s(1 - \bar{p}_{+x})^{f_+ - s},
    \label{eq:Cs}
\end{equation}

\noindent which is simply a binomial distribution for the solvation environments. Hence, the most common solvation environment will be the mode of the binomial distribution with parameters $f_+$ and $\bar{p}_{+x}$, explicitly shown in the SI. The values of $\bar{p}_{+x/y}$ are computed from MD simulations (using the ensemble average coordination numbers) and used in the theory, which can also be used to calculate the ratio of the association constants $\bar{\lambda}_x/\bar{\lambda}_y$, using the conservation of associations. In Fig.~\ref{fig:sticky_helm}b) we show the theory for the $-0.4$~e~nm$^{-2}$ case, which clearly agrees well with the MD simulations.



The MD results for the $-0.6$~e~nm$^{-2}$ are shown in Fig.~\ref{fig:sticky_helm}c). We find that the most probably solvation environment is EC$+$2EMC, with 2EC$+$EMC and 3EC also be possible, but practically no other solvation environment. Therefore, for this surface charge, a better functionality would be $f_+ = 3$. In Fig.~\ref{fig:sticky_helm}d) we show the corresponding theory plot using $f_+ = 3$, which agrees reasonably well with the MD simulations. The most probable solvation environment is 2EC$+$EMC, but the EC$+$2EMC is a similar probability.



Finally, for the most negative surface charge results for MD simulations can be found in Fig.~\ref{fig:sticky_helm}e). Here we find the most probable solvation environment to be 2EC$+$EMC, with the next most likely being 3EMC. Again a functionality of $f_+ = 3$ appears to be a natural choice. In Fig.~\ref{fig:sticky_helm}f) we show the corresponding theory plot, which predicts EC$+$2EMC to be the most likely, with 2EC$+$EMC to be the next most likely. 

Overall, the agreement is reasonable between the theory and MD simulations, and these results demonstrate that a reduced functionality works well to describe the solvation environments in the Helmholtz layer. This is perhaps not surprising, as the Li$^+$ will interact strongly with a charged interface, and block at least one association site of Li$^+$. 
Therefore, when constructing a theory for the Helmholtz layer, we must not use the same functionality in all space, but must reduce it at the interface, meaning that $f_+$ also becomes an EDL quantity. In the SI, we more explicitly demonstrate that using $f_+ = 5$, as in the bulk, and only changing the association constant does not provide a satisfactory match with the MD simulations. 

In Tab.~\ref{tab:HLQ} we display the ratio of the association constants, $\lambda_x/\lambda_y$, and the molar ratio of the solvents. In contrast to the diffuse EDL, we find $\lambda_x/\lambda_y$ is reduced by more than an order of magnitude at the interface. Note that $\lambda_x/\lambda_y$ does explicitly depend on $f_+$, but only weakly so through the mass action laws. Therefore, this large reduction is not expected from this change small change in $f_+$, but we anticipate it is from another source. It can also be seen that the molar ratio of EC is much larger than the bulk, but it appears to saturate near 4$\times$. 

As found by Markiewitz \textit{et al.}~\cite{Markiewitz25WiSE} for WiSE (from theory and MD simulations), the $\lambda_i$ between Li$^+$ and solvents can vary in the EDL if the solvents have a significant dipole moment and can be described as a fluctuating Langevin dipole. This arrives from assumption that the solvent only behaves as a fluctuating dipole when it is in the free state, so not bound to a Li. Therefore, as the dipole moment of EC is much larger than EMC, we would expect $\lambda_x/\lambda_y$ to decrease with increasing electric field, and for the amount of EC to increase relative to EMC as it has a large dipole moment, it will be energetically favorable for it to reside in the larger electric fields. These observations are included in a new theory of the Helmholtz layer, which is outlined in the SI in detail. In the following section, we present a simplified analysis of this theory.

\subsection{Helmholtz Layer Solvation from Bulk Solvation}


In the SI we outline in full the new theory for the solvation in the Helmholtz layer. To illustrate its important points, we will solve a back-of-the-envelope example here, not solving the system of equations in its full complexity. Our aim is to demonstrate some of its trends, without getting into details too much. We will assume that there are no anions in the Helmholtz layer (observed in MD for moderate negative surface charges), and that the volume fraction of Li$^+$ cations is constant (also observed in MD, at least approximately) and we assume the volumes of each solvent are identical, which means that the only changes occurring is from the solvents swapping places. Note we treat the solvents as fluctuating Langevin dipoles when they are free, but not when they are bound to Li$^+$. Therefore, the equations which need to be solved only depend on electric field, which we can approximate from the surface charge density of the simulations. 




Furthermore, we will work with the sticky-cation approximation, such that the solvent distributions are described by Eq.~\eqref{eq:Cs}. The association probabilities for which can be calculated from 

\begin{align}
    &\psi_+ p_{+x} = \psi_x p_{x+}=\frac{\psi_y-\psi_++\lambda(\psi_++\psi_x)}{2(\lambda-1)} \nonumber \\
    &-\frac{\sqrt{4\psi_y\psi_+(\lambda-1)+[\lambda(\psi_x-\psi_+)+\psi_++\psi_y]^2}}{2(\lambda-1)},
    \label{eq:prob_1_sticky}
\end{align}
\noindent and
\begin{align}
    &\psi_+ p_{+y} = \psi_y p_{y+} =\frac{\psi_y+\psi_++\lambda(\psi_x-\psi_+)}{2(1-\lambda)} \nonumber \\
    &-\frac{\sqrt{4\psi_y\psi_+(\lambda-1)+[\lambda(\psi_x-\psi_+)+\psi_++\psi_y]^2}}{2(1-\lambda)},
    \label{eq:prob_2_sticky}
\end{align}

\noindent which is the solution of the mass action laws for just solvent in the sticky-cation case. Note that a bar is used to denote quantities within the EDL/Helmholtz layer ($\bar{\psi}_i$,$\bar{\lambda}$,etc.), which are omitted from Eqs.~\eqref{eq:prob_1_sticky}-\eqref{eq:prob_2_sticky} for clarity. Here, the ratio of the association constants in the Helmholtz layer is given by 
\begin{equation}
\bar{\lambda} = \dfrac{\lambda_x}{\lambda_y}\dfrac{p_x}{p_y}\dfrac{\sinh(\beta p_y|\nabla \Phi|)}{ \sinh(\beta p_x|\nabla \Phi|)},
\label{eq:barL}
\end{equation}

\noindent where $p_x$ and $p_y$ are the dipole moments of EC and EMC, respectively, and $\Phi$ is the electrostatic potential, with $-\nabla \Phi$ being the electric field. As $p_x > p_y$, the ratio of the association constants decreases with increasing electric field, which is a reflection of EC gaining energy as being a free fluctuating dipole. This was observed previously, as seen in Tab.~\ref{tab:HLQ}.

Next, to determine the composition in the Helmholtz layer, we need to know the volume fractions of each solvent. This can be obtained from a Boltzmann closure relation of the solvents and a statement of incompressibility, following Goodwin \textit{et al.}~\cite{Goodwin2022GEL}. Here, we assume the Boltzmann closure takes the form
\begin{equation}
    \dfrac{\bar{\phi}_{0010}}{\bar{\phi}_{0001}} = \dfrac{\phi_{0010}}{\phi_{0001}}\dfrac{p_y}{p_x}\dfrac{ \sinh(\beta p_x|\nabla \Phi|)}{\sinh(\beta p_y|\nabla \Phi|)}.
    \label{eq:BCR_sq}
\end{equation}

\noindent In the SI, the full set of closure relations are shown, with the surface interaction terms and Lagrange multiplier for asymmetric sizes, but we only investigate the simplified form here. We also take $\phi_x + \phi_y = \phi_{xy}$, where $\phi_{xy} < 1$ is the constant volume fraction of solvent. 

From substituting Eq.~\eqref{eq:barL} into the Boltzmann closure relation, we can simplify Eq.~\eqref{eq:BCR_sq} to obtain
\begin{equation}
    \dfrac{\bar{\phi}_x\bar{p}_{x+}}{\bar{\phi}_y\bar{p}_{y+}} =  \dfrac{\phi_xp_{x+}}{\phi_yp_{y+}} ,
    \label{eq:cons}
\end{equation}

\noindent which can also be stated as 
\begin{equation}
    \dfrac{\bar{p}_{+x}}{\bar{p}_{+y}} =  \dfrac{p_{+x}}{p_{+y}},
    \label{eq:bulk_H}
\end{equation}

\noindent and therefore, \textit{this approximation states that the probability that the solvents are binding to the association sites do not change from the bulk solvation probabilities}. The bulk value computed for $p_{+x} \approx 0.48$, and the values for $\bar{p}_{+x}$ are shown in Tab.~\ref{tab:HLQ}, which can be seen to be close to the bulk value. Therefore, the MD simulations appear to approximately follow this conservation of solvation probabilities. Reflecting on the solvation distributions in the Helmholtz layer, and also the diffuse EDL, it can be seen that they do not qualitatively change with surface charge, with the only significant change being the change in functionality, which further supports the simple theory findings here. 



As the functionality is reduced in the Helmholtz layer, the numbers of coordinated solvent still decrease, but their relative population in the solvation shell does not change. While the $p_{+x/y}$ does not change, at least given the approximations here, Eq.~\eqref{eq:cons} does not state that $\phi_{x/y}$ and $p_{x/y+}$ need to stay constant, but that the ratio of the $\Gamma_{x/y}$ values remains the same as the bulk. In fact, we know the volume fractions significantly change, as seen from the large enhancement of free EC in Eq.~\eqref{eq:BCR_sq}, and $p_{x/y+}$ must change because of this, but this is compensated in the change in Eq.~\eqref{eq:barL}.





The volume fractions of solvents and $p_{x/y+}$ could be obtained from Eq.~\eqref{eq:cons} and Eqs.~\eqref{eq:prob_1_sticky}-\eqref{eq:prob_2_sticky}, while using the incompressibility constraint ($1 = \phi_+ + \phi_x + \phi_y$). To solve this system of equations, we use a $\phi_+ = 0.015$ (approximately what we find in MD, using volume ratios in Ref.~\citenum{Goodwin2023PRXE}), $\lambda_x/\lambda_y = 3.7$ (found from the bulk MD section), and for the dipole moments $p_x = 5$~D~\cite{Payne9172} and $p_y = 1$~D~\cite{Thomson1939}. Using a dielectric constant of 5, we find $\bar{\lambda}_{-0.4} = 0.277$, $\bar{\lambda}_{-0.6} = 0.048$ and $\bar{\lambda}_{-0.8} = 0.008$. From solving the system of equations with these parameters, we find $\bar{x}_x/\bar{x}_y|_{-0.4} = 2.42$ while using $f_+ = 4$, and $\bar{x}_x/\bar{x}_y|_{-0.6} = 4.98$ and $\bar{x}_x/\bar{x}_y|_{-0.8} = 5.84$ from using  $f_+ = 3$. This demonstrates that even though the solvation distribution of Li$^+$ is not significantly changing in the Helmholtz layer, mainly through the reduced functionality, the amounts of each solvent are significantly changing. While the agreement is not exact against MD simulations, as seen in Tab.~\ref{tab:HLQ}, the qualitative agreement is reasonable. 








\subsection{Additional Electrolytes}

%
%

In the SI, we further tested the other electrolytes investigated in Ref.~\citenum{Wu2023}. Specifically, the ether solvent mixture with 1,3-dioxolane (DOL) and  1,2-dimethoxyethane(DME), was investigated with lithium bis(trifluoromethanesulfonyl)imide (LiTFSI). The ether-based solvents typically interact with the Li$^+$ less strongly than the carbonate-based electrolytes, which makes the comparison to our theory more challenging. We find that a functionality of 4 is more appropriate here, but that when focusing on only the solvation environments (without ionic aggregate), a functionality of 3 might fit the data better. Despite this more difficult electrolyte, we still observe similar trends to the case of EC-EMC in the main text. Specifically, that the dominant effect in the Helmholtz layer is the apparent reduction in the functionality of Li$^+$. However, we only find that $p_{+x}$ remains (approximately) constant at moderate surface charges, with significant deviations from the bulk value being observed for large surface charges. This demonstrates that the assumptions the result in Eq.~\eqref{eq:bulk_H} are not universal, and that while the bulk solvation environments are a good starting point, the solvation environments in the EDL should still be investigated.

For the LiTFSI-DOL+DME case, the cation-anion interactions are typically more pronounced than the LiPF$_6$ interactions. In the diffuse layer of this electrolyte, we further investigated the ionic aggregation effects, which is shown in the SI. At moderate surface charges (0.6 enm$^{-2}$), we find that there are some aggregates larger than ion pairs, which typically does not occur for the other studied surface charges. This suggests that ionic aggregation could be enhanced at some moderate voltages, which was previously found by Markiewitz \textit{et al.} in the context of salt-in-ionic liquids~\cite{Markiewitz24,Zhang24} and water-in-salt electrolytes~\cite{Markiewitz25WiSE}. 

Moreover, in Ref.~\citenum{Wu2023}, the solvation in the EDL with the additive fluoroethylene carbonate (FEC) was investigated. In the SI, we also investigated this three solvent case. We found that under analogous assumptions to Eq.~\eqref{eq:Cs}, the same holds for the three-solvent case, and therefore, it generalizes to any number of solvents, given analogous assumptions. Overall, we find the EC+EMC+FEC cases behaves similar to the EC+EMC mixture, and DOL+DME+FEC behaves in an analogous was to DOL+DME, and therefore, we will not discuss these cases further here. 

\section{Discussion}


%
%
Overall, the main effect we observe from thoroughly analyzing the solvation environments in the Helmholtz layer of non-aqueous battery electrolytes is that they (largely) appear to be the same as the bulk solvation environments, but where the number of association sites of Li$^+$ is reduced. This is perhaps not surprising, as the interface physically blocks some association sites and interacts with the Li$^+$. Moreover, reduced coordination numbers of solvents has been reported in myriad other simulations of non-aqueous battery electrolytes~\cite{von2012correlating,Wu2023,Oyakhire2023,Wu25}. However, here we show that the effect is best described through changing the number of available binding sites, instead of only changing the Li-solvent interactions. 

There are several implications of this observation. As an equilibrium between the Helmholtz layer and bulk must be established, it becomes apparent that even without any applied fields or interactions with the surface, \textit{the electrolyte can become charged from the reduction of functionalities at the interface}. Moreover, it does not appear that it is necessary to establish an equilibrium between the diffuse EDL and the Helmholtz layer, although one could be established, but as both of them are in equilibrium with the bulk, they should both be in equilibrium with each other. 

%
%

Under certain assumptions, we found that the \textit{cation-solvent association probabilities remain constant in the Helmholtz layer, and moreover, equal to their bulk values}. If these assumptions apply to an electrolyte, it means only the bulk solvation environments need to be investigated, and the Helmholtz layer solvation environments can simply be predicted from the developed theory with a reduced $f_+$. We found that EC+EMC approximately follows these assumptions, but that DOL+DME did only for small surface charges. Therefore, this observation is not universal, and the assumptions can be broken in real electrolyte systems, which means investigating the EDL of these electrolytes is still necessary to test if this observation holds. Moreover, we found that sometimes the functionality reduces by 1, but it can reduce by 2, and also depend on surface charge. Therefore, performing EDL simulations of the electrolytes is still important to establish their functionalities in the Hemholtz layer.

%
%

The theory developed here is a simple lattice-gas mean-field theory, that accounts for correlations beyond mean-field through the associations between species. While it is not sophisticated, it is analytically tractable and physically interpretable. However, it does certainly miss some correlations beyond mean-field and struggles with the spatial resolution of species in the EDL, as does any local density approximation. For further discussion of the limitations of the theory see Refs.~\citenum{Goodwin2022GEL,Markiewitz24}. Note that previously we identified the solvation/ionic aggregation effects right at the interface as a limitation of the formalism developed by Goodwin and Markiewitz \textit{et al.}, but here we have at least shown for non-aqueous Li-ion battery electrolytes it can provide insight. 

%
%

Currently, there doesn't appear to be a standardized convention for reporting solvation environments and ionic aggregates, other than showing coordination numbers. As discussed in Ref.~\citenum{Goodwin2023PRXE}, coordination numbers do not provide a unique classification of the associations, and further information is required, such as the cluster bond density. Here we have found it insightful to plot the solvation distributions for free cations, as a function of the number of each bound solvent. This provided a natural way to visualize the results, which provided insight into the Helmholtz layer. Therefore, we again suggest that the reporting convention outlined in Ref.~\citenum{Goodwin2023PRXE} can provide a natural framework to work within to further understand these complex electrolytes.

%
%

Looking forward, the formalism developed here could be extended to a Stern model, where both Helmholtz layer and diffuse EDL are combined in series, and perhaps integrated further. Moreover, the theory could be integrated with microscopic models of electrochemical reaction kinetics, such as coupled ion-electron transfer theory~\cite{bazant2023unified}, to theoretically investigate possible reactions at interfaces. Finally, the motivation of studying solvation environments in the first monolayer of an electrified interface, i.e. the Helmholtz layer, is to predict the species that may reacting at the interface to form the SEI. Therefore, our theory could be used to predict possible solvation environments in the Helmholtz layer from bulk solvation environments from MD simulations or experiments, and then used in DFT to see the reductive stability of these solvation environments, or using them as inputs/biases for reaction networks to predict what could form in the SEI.








\section*{Acknowledgments}

We are grateful to J. Pedro de Souza and M McEldrew for the helpful discussions. Z.A.H.G acknowledges support through the Glasstone Research Fellowship in Materials and The Queen's College, University of Oxford. D.M.M. \& M.Z.B. acknowledge support from the Center for Enhanced Nanofluidic Transport 2 (CENT$^2$), an Energy Frontier Research Center funded by the U.S. Department of Energy (DOE), Office of Science, Basic Energy Sciences (BES), under award \# DE-SC0019112. D.M.M. also acknowledges support from the National Science Foundation Graduate Research Fellowship under Grant No. 2141064. Q.W. and Y.Q. acknowledge NASA for financial support (grant no. 80NSSC21M0107). 





\bibliography{main.bib}

\end{document}